\documentclass[reprint, aps, prd,twocolumn, nofootinbib]{revtex4}  %

\usepackage{graphicx}%
\usepackage{dcolumn}%
\usepackage{bm}%
\usepackage[linktocpage,colorlinks,pdfusetitle,allcolors=blue]{hyperref}
\usepackage{amsmath, amssymb}
\usepackage{siunitx}

\renewcommand{\d}[1]{\ensuremath{\operatorname{d}\!{#1}}}

\begin{document}

\title{Can quasi-circular mergers of charged black holes produce extremal black holes?}

\author{Gabriele Bozzola}
\email{gabrielebozzola@arizona.edu}
\affiliation{Department of Astronomy, University of Arizona, Tucson, AZ, USA}

\author{Vasileios Paschalidis}
\email{vpaschal@arizona.edu}
\affiliation{Department of Astronomy, University of Arizona, Tucson, AZ, USA}
\affiliation{Department of Physics, University of Arizona, Tucson, AZ, USA}

\date{\today}

\begin{abstract}
  In contrast to energy and angular momentum, electric charge is conserved in
  mergers of charged black holes. This opens up the possibility for the remnant
  to have Kerr-Newman parameter $\chi^{2} + \lambda^{2}$ greater than 1 (with $\chi$ and $\lambda$
  being the black hole dimensionless spin and dimensionless charge,
  respectively), which is forbidden by the cosmic censorship conjecture. In this
  paper, we investigate whether a naked singularity can form in quasi-circular
  mergers of charged binary black holes. We extend a theoretical model to
  estimate the final properties of the remnant left by quasicircular mergers of
  binary black holes to the charged case. We validate the model with
  numerical-relativity simulations, finding agreement at the percent level. We
  then use our theoretical model to argue that while naked singularities cannot
  form following quasi-circular mergers of non-spinning charged binary black
  holes, it is possible to produce remnants that are arbitrarily close to the
  extremal limit.
\end{abstract}

\maketitle

\section{Introduction}%
\label{sec:introduction}

This paper is concerned with quasi-circular mergers of electrically charged
black holes. To set the stage of our work, it is convenient to first consider
the more familiar case without charge. In 2006, it was observed that the latest
stages of the inspiral of two highly spinning black holes are significantly
different compared to the non-spinning counterpart~\cite{Campanelli2006}. The
main difference is that inspiral of spinning black holes takes substantially
longer than the non-spinning case -- an effect that is referred to as the
\emph{orbital hang-up}. One way to understand why this happens involves
conservation of angular momentum and the cosmic censorship conjecture. If all
the angular momentum available in the system (orbital + spins) were to end up in
the remnant, the object would be over-extremal, i.e.\ its dimensionless spin
$\chi$ would be larger than 1. Such a black hole is not possible in general
relativity, and Kerr spacetimes with $\chi>1$ are not black holes, but naked
singularities (see, e.g.~\cite{Wald1984}). Given that the formation of naked
singularities is forbidden by the cosmic censorship conjecture,\footnote{Note,
however that the formation of naked singularities in fine-tuned dynamical
scenarios is possible~\cite{Choptuik1993}} the binary has to radiate away all
the excess angular momentum to be able to merge. To do so, the black holes
inspiral for longer so that the gravitational waves can carry away the excess
angular momentum.

Now, consider mergers of charged black holes. While energy and angular momentum
can be radiated away, electric charge is always conserved. For this reason, a
natural question to ask is whether it is possible to start from charged black
holes with individual charge-to-mass ratio $\lambda < 1$ and form an extremal
remnant.\footnote{In~\cite{Bozzola2022}, we investigated a similar question
  checking whether ultra-relativistic head-on collisions of black holes can lead
  to extremal configurations. We found that there is no indication that this can
  happen. In that case, the formation of a naked singularity was avoided by the
  large kinetic energy in the system.} If this does not happen, how is the
formation of a naked singularity avoided? Is there a charge-induced orbital
hang-up? Does the system inspiral or does it outspiral after sufficient energy
has been radiated away? This paper aims to answer these questions by extending
the method described in~\cite{Buonanno2008, Kesden2008} to charged black hole
binaries. We are going to refer to this method as ``BKL''\footnote{Not to be
  confused with the BKL singularity studied by Belinski–Khalatnikov–Lifshitz.}
from the initials of the original authors~\cite{Buonanno2008}.\footnote{The
  contribution of~\cite{Kesden2008} is to include the loss of energy due to
  gravitational wave. As we will discuss later, this is needed to match the to
  reach percent-level agreement with the numerical relativity simulations.} The
approach is based on conservation arguments and analogy with point particles.
Previous studies have shown that this simple argument is surprisingly effective
at capturing the remnant properties to within a percent~\cite{Kesden2008}. We
validate our extended model via numerical relativity simulations of
quasi-circular mergers of non-spinning, charged binary black holes, and use it
to argue that quasi-circular inspirals of charged binary black holes cannot form
naked singularities. We focus on configurations in which the black holes have
charge with the same sign so that the total charge is greater than the
individual black hole charges, which give the remnant black hole the possibility
to maximize the Kerr-Newman parameter through its charge.

The goal of our paper is to explore and understand better cosmic censorship in
the non-linear regime and see if it introduces novel effects. Several linear
arguments argued that black holes cannot be overcharged~\cite{Wald2018}, but the
question is still open for the non-linear case. Moreover, while this paper
focuses on fundamental physics implications, the analytic model predicting the
remnant properties that is developed here has direct astrophysical
applications. Charge in the astrophysical context of binary mergers has recently
received some attention as ``charge'' can acquire different meanings from
magnetic monopoles to modified gravity (see, e.g., Introduction
in~\cite{Bozzola2019}). Additionally, remnant properties can be useful when
studying populations of magnetically charged primordial black holes (see,
e.g.,~\cite{Liu2020}) and in gravitational-wave astronomy, where they are the
starting point to study the ringdown signal (see, e.g.~\cite{Carullo2021}). The
analytic model described in this paper is computationally efficient, which makes
it optimally suited for quick estimates.

This paper is structured as follows. In Section~\ref{sec:setup}, we describe the
formalism we developed for predicting the properties of the remnant of charged
binary black holes. In Section~\ref{sec:results}, we show and discuss our
results. We conclude with Section~\ref{sec:conclusions}. We work in geometrized
and gaussian units with $G = c = {(4\pi\varepsilon_{0})}^{-1} = 1$, with $G$
being Newton's constant, $c$ the speed of light in vacuum, and $\varepsilon_{0}$
the permittivity of vacuum.

\section{Setup}%
\label{sec:setup}

We approach the problem of quasi-circular mergers of charged black holes with an
analytical model that we validated with numerical relativity simulations. In
Section~\ref{sec:analytical-model}, we discuss the analytical method, and in
Section~\ref{sec:numer-simul} we present our framework for the full non-linear
calculations.

\subsection{Analytical model}%
\label{sec:analytical-model}

To estimate the properties of the remnant left by the merger of two charged
black holes, we follow Ref.~\cite{Kesden2008}, which extended the approach
outlined of Ref.~\cite{Buonanno2008}. This method is based on conservation
principles and an effective one-body treatment. The core assumption of the model
is that energy and angular momentum lost can be determined by looking at the
properties of the innermost stable circular orbit (ISCO) of a properly computed
effective background spacetime~\cite{Hughes2003,Buonanno2008,Kesden2008}. This
is because the system loses the vast majority of its initial energy and angular
momentum during the inspiral, and when it reaches the ISCO, the plunge is so
rapid that there is no significant loss of energy and angular momentum (i.e.,
the emission is small compared to the rest of the inspiral). In a nutshell, the
method consists of finding a suitable background spacetime and computing the
properties of its innermost-stable circular orbit.

\subsubsection{The effective-one-body problem}%
\label{sec:effective-one-body}

BKL~\cite{Buonanno2008} propose to treat the general relativistic two body
problem as if it was in the limit of extreme mass ratio, where the system is
equivalent to a test mass in a background spacetime. Strictly speaking, this
approach is invalid for comparable masses, but previous work found that the
agreement with the full non-linear solution is
excellent~\cite{Buonanno2008,Kesden2008}. Hence, we adopt the same basic idea
and extend it to include the charge in black hole spacetimes.

Consider two black holes that are separated by a distance that is large enough
so that we can give them a well-defined mass and charge $m_{1}$, $m_{2}$, and
$q_{1}$, $q_{2}$, with total mass and charge $M = m_{1} + m_{2}$, $Q = q_{1} +
q_{2}$.  The equivalent one-body problem has a test-mass with mass
$m_{\text{red}}$, charge $q_{\text{red}}$ in a Kerr-Newman spacetime with mass
$M$, charge $Q$, and angular momentum $J$. Here, $m_{\text{red}}$ and
$q_{\text{red}}$ are the reduced mass and charge, defined as
\begin{subequations}%
  \begin{align}
    m_{\text{red}} &= \frac{m_{1} m_{2}}{M}\,, \\
    q_{\text{red}} &= \frac{q_{1} q_{2}}{Q}\,.%
  \end{align}%
\label{eq:reduc}%
\end{subequations}%
As we will see later, it is more convenient to work with dimensionless charge
$\lambda = Q \slash M$, $\textsf{q} = q_{\text{red}} \slash m_{\text{red}}$, and dimensionless
spin $\chi = J \slash M^{2}$.

In the BKL framework, the energy (angular momentum) radiated by gravitational
waves is the orbital energy (angular momentum) of the test particle up to the
innermost-stable circular orbit\@.\footnote{Note that there are two ISCOs,
  prograde and retrograde with respect to the rotation of the black hole. In
  this paper, we only consider prograde ones. } We make the same assumption and
compute final mass, spin, and charge for a binary merger by studying the ISCO\@.
If $\varepsilon^{\textsf{q}}_{\text{ISCO}}(\lambda, \chi)$ is the specific energy at the ISCO for a
particle with reduced charge \textsf{q} in a Kerr-Newman spacetime with
charge-to-mass ratio $\lambda_{\text{final}}$ and dimensionless spin
$\chi_{\text{final}}$ (note, $\varepsilon$ is independent of the mass $M$), the energy
radiated is
\begin{equation}
  \label{eq:1}
  E_{\text{GW}} = m_{\text{red}} - m_{\text{red}}\varepsilon^{\textsf{q}}_{\text{ISCO}}(\lambda_{\text{final}}, \chi_{\text{final}}),
\end{equation}
where $\lambda_{\text{final}}$ and $\chi_{\text{final}}$ have yet to be determined.
Conservation of energy implies that
\begin{equation}
  \label{eq:mass-final}
  M_{\text{final}} = M - E_{\text{rad}} = M(1 - \nu(1 - \varepsilon^{\mathsf{q}}_{\text{ISCO}}(\chi_{\text{final}}, \lambda_{\text{final}})))\,,
\end{equation}
where $E_{\text{rad}}$ is the total energy emitted in gravitational and
electromagnetic waves, and $\nu = m_{\text{red}} \slash M$ is the symmetric mass
ratio. Similarly, all the angular momentum is radiated away except for the
amount available at the ISCO, which is $J_{\text{ISCO}} = m_{\text{red}} M
l^{\textsf{q}}(\lambda_{\text{final}}, \chi_{\text{final}})$, where $\ell$ is
the dimensionless angular momentum at the ISCO of a Kerr-Newman spacetime with
charge $\lambda_{\text{final}}$ and spin $\chi_{\text{final}}$.\footnote{$M\ell$
is the specific angular momentum, so $m_{\text{red}} M \ell$ is the actual
angular momentum.} Therefore, we have that
\begin{equation}
  \label{eq:chi-final}
  \chi_{\text{final}} = \frac {J_{\text{ISCO}}}{{M_{\text{final}}}^{2}} = \frac{\nu \ell_{\text{ISCO}}(\chi_{\text{final}}, \lambda_{\text{final}})}{{{\left[1 - \nu(1 - \varepsilon_{\text{ISCO}}(\chi_{\text{final}}, \lambda_{\text{final}}))\right]}}^{2}}\,.
\end{equation}
Charge is conserved, so $Q_{\text{final}} = Q = q_{1} + q_{2}$, and
\begin{equation}
  \label{eq:lambda-final}
  \lambda_{\text{final}} = \frac {Q}{{M_{\text{final}}}} =
  \frac{\lambda}{1 - \nu(1 - \varepsilon_{\text{ISCO}}(\chi_{\rm final}, \lambda_{\rm final}))}\,.
\end{equation}
Finally, we have the coupled system of non-linear algebraic equations
\begin{subequations}%
  \label{eq:final_system}
  \begin{align}
    \chi_{\text{final}} &= \frac{\nu \ell^{\textsf{q}}_{\text{ISCO}}(\chi_{\text{final}}, \lambda_{\text{final}})}{{[1 - \nu(1 - \varepsilon^{\textsf{q}}_{\text{ISCO}}(\chi_{\text{final}}, \lambda_{\text{final}}))]}^{2}}\,, \\
    \lambda_{\text{final}} &= \frac{\lambda}{1 - \nu(1 - \varepsilon^{\textsf{q}}_{\text{ISCO}}(\chi_{\text{final}}, \lambda_{\text{final}}))}\,.
  \end{align}%
\label{eq:kesden}%
\end{subequations}
The unknowns in these equations are $\chi_{\text{ISCO}}$ and $\lambda_{\text{ISCO}}$,
that we find numerically with the Levenberg-Marquardt
algorithm~\cite{Levenberg1944,Marquardt1963} as implemented in the \texttt{root}
function in \texttt{scipy.optimize}~\cite{SciPy}.

\subsubsection{ISCO for a charged particle in a Kerr-Newman spacetime}%
\label{sec:isco-kerr-newman}

To solve the system defined by Eqs.~\eqref{eq:final_system}, we need to compute
the dimensionless energy $\varepsilon$ and angular momentum $\ell$ for particles with
charge-to-mass ratio $\mathsf{q}$ on the ISCO of Kerr-Newman spacetimes. If we
focus on the equatorial plane and work in Boyer-Lindquist coordinates
$(t,r,\theta,\phi)$, these quantities can be calculated using an effective potential
$V_{\text{eff}}(r)$. For Kerr-Newman black holes with unit mass,\footnote{Note,
  $\varepsilon$ and $\ell$ are independent of the mass.} charge $\lambda$, and spin $\chi$,
$V_{\text{eff}}$ is given by~\cite{Jaiakson2017,Siahaan2020, Bozzola2021} (see,
Section IV A.\ in~\cite{Bozzola2021})
\begin{multline}
  \label{eq:Veff}
  V_{\text{eff}}(r) = \frac{1}{r^{4}} \left[ - \Delta(r)
    + (\Delta(r) - \chi^{2})\tilde{l}^{2}(r) \right. \\
    \qquad - 2 \chi (r^{2} + \chi^{2} - \Delta(r)) \tilde{l}(r)\tilde{\varepsilon}(r) \\
  \left. + \left( {(r^{2} + \chi^{2})}^{2} - \Delta(r) \chi^{2} \right) \varepsilon^{2}(r)\right] \,,
\end{multline}
with
\begin{subequations}
  \begin{align}
    \Delta(r)   & = r^2 - 2r + \chi^2 + \lambda^2\,, \\
    \tilde{l}(r) &= l(r) + q \frac{\lambda}{r} \chi\,, \\
    \tilde{\varepsilon}(r) &= \varepsilon(r) + q \frac{\lambda}{r}\,,
  \end{align}
\end{subequations}
where $\varepsilon(r)$ and $l(r)$ are the specific energy and dimensionless angular
momentum ($\ell = a \slash m_{\text{red}}$) for circular orbits of radius $r$. The
properties of the ISCO are found solving the following equations simultaneously
\begin{subequations}%
  \label{eq:isco}
  \begin{align}
    V_{\text{eff}}(r_{\text{ISCO}}) &= 0\,, \label{eq:isco1} \\
    \frac{\d V_{\text{eff}}}{\d r}(r_{\text{ISCO}}) &= 0\,,  \label{eq:isco2} \\
    \frac{\d{}^{2} V_{\text{eff}}}{\d r^{2}}(r_{\text{ISCO}}) &= 0\,, \label{eq:isco3}
  \end{align}
\end{subequations}
for $r_{\text{ISCO}}$, $\varepsilon_{\text{ISCO}} = \varepsilon(r_{\text{ISCO}})$ and
$\ell_{\text{ISCO}} = \ell(r_{\text{ISCO}})$. Eq.~\eqref{eq:isco1} and
Eq.~\eqref{eq:isco2} impose circularity of the orbit, Eq.~\eqref{eq:isco3} the
condition of being innermost stable.

These equations can be solved analytically, but it is simpler and faster to
solve them numerically. In practice, we use \texttt{SymPy}~\cite{SymPy} to
derive symbolically $V_{\text{eff}}(r)$ and
Eqs.~\eqref{eq:isco}. Then, we solve this system numerically
using the \texttt{root} function in \texttt{scipy.optimize}~\cite{SciPy}. This
gives us $\varepsilon_{\text{ISCO}}^{\mathsf{q}}(\lambda, \chi)$ and
$l_{\text{ISCO}}^{\mathsf{q}}(\lambda, \chi)$ for a particle of charge-to-mass
ratio $\mathsf{q}$ in a Kerr-Newman spacetime with dimensionless charge and spin
$\lambda$ and $\chi$. This is what we need in order to solve
Eqs.~\eqref{eq:final_system}. The Python code that implements the entire scheme
is provided in the Supplemental Material.

\subsection{Numerical simulations}%
\label{sec:numer-simul}

We validate the model described in the previous section using two sets of
numerical relativity simulations. First, we use the simulations of the
quasi-circular inspiral and merger of unequal-mass (mass ratio of 29/36),
charged black holes we presented in~\cite{Bozzola2020, Bozzola2021, Bozzola2022,
  Luna:2022udb}. Second, we perform new simulations with higher resolution and
charge. The second set consists of eleven simulations with charge-to-mass ratio
up to $\lambda = 0.6$ (like sign charge) and equal mass. Systems with higher $\lambda$ take
a significantly longer time to merge and are much more computationally
expensive.

Our numerical relativity simulations solve the coupled Einstein-Maxwell
equations in a $3+1$ decomposition of the spacetime (for more details,
see,~\cite{Alcubierre:2008it,Baumgarte:2010nu,Shibata2016b}) and use the
\texttt{Einstein
  Toolkit}~\cite{Loffler:2011ay,EinsteinToolkit:ascl,EinsteinToolkit:web,EinsteinToolkit:2021_11}
for the numerical integration. We generate initial data with
\texttt{TwoChargedPunctures}~\cite{Bozzola2019} for systems of two black holes
with fixed charge-to-mass ratio $\lambda$. We use sixth-order finite-difference
methods to evolve the spacetime with the \texttt{Lean} code~\cite{Sperhake2007},
which implements the Baumgarte-Shapiro-Shibata-Nakamura (BSSN) formulation of
Einstein's equations~\cite{Shibata1995,Baumgarte1998} and the electromagnetic
fields are evolved with the massless version of the
\texttt{ProcaEvolve}~\cite{Zilhao2015} code, part of Canuda
suite~\cite{canuda,canudacode}. We locate apparent horizons with
\texttt{AHFinderDirect}~\cite{Thornburg:1995cp,Thornburg:2003sf}, and their
physical properties are measured with \texttt{QuasiLocalMeasuresEM}, a version
of \texttt{QuasiLocalMeasures}~\cite{Dreyer:2002mx} updated to implement the
isolated horizon formalism in full Einstein-Maxwell theory (see Sec.~II~C
in~\cite{Bozzola2019}).

We work with Cartesian grids with Berger-Oliger adaptive mesh refinement as
provided by \texttt{Carpet}~\cite{Schnetter:2003rb}. The simulations use between
nine and thirteen refinement levels centered on and tracking the centroid of the
black hole apparent horizons. The initial separation is \num{12.1}\,$M$, where
$M$ is the total ADM mass of the system. Eccentricity is kept below \num{0.01}
with the method described in~\cite{Bozzola2021}. For the higher values of
charge, we also had to manually adjust the initial momenta to meet the target
eccentricity. We did so through trial and error. The resolution of our
unequal-mass simulations is $M\slash65$ for the unequal mass case, whereas out set of
equal-mass simulations has charge-dependent resolution as follows: the finest
grid spacing set to
$\Delta x_{\text{finest}} = \sqrt{1 - \lambda^{2}}\slash 320\,M$,\footnote{In isotropic
  coordinates, the horizon radius for a Reissner-Nordstr{\"o}m black hole with
  mass $M_{1} = \num{0.5}\,M$ and charge $Q_{1} = \lambda M_{1}$ is
  $\sqrt{1 - \lambda^{2}} \slash 4\,M$.} which ensures that the horizons are resolved with
more than 80 grid points. The damping parameters $\eta$ and $\kappa$ in the evolution
equations of the shift vector and the electric field were set to $1.5\, M$ and
$10\, M$. We refer the reader to~\cite{Bozzola2020, Bozzola2021, Bozzola2022}
for a detailed complete discussion of the methods and tools we employ.

In addition to the convergence studies described in~\cite{Bozzola2020,
  Bozzola2021, Bozzola2022}, we performed more simulations at higher resolution
to estimate errors an convergence properties. In all cases, we find that the
quasi-local properties of the black hole (mass, spin, charge) are exceptionally
well-behaved and we estimate the numerical error due to finite resolution to be
at the level of \SI{0.1}{\percent}.

\section{Results}%
\label{sec:results}

\begin{figure}[htbp]
  \centering
  \includegraphics[width=\linewidth]{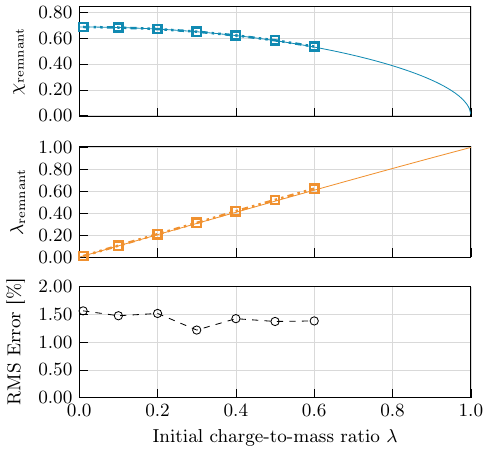}
  \caption{Top two panels: Physical properties of the remnants left by the
    merger of two equal-mass, equal-charge black holes for the
    numerical-relativity simulations (squares) and the analytical predictions
    (solid lines). Bottom panel: Total relative error, measured as in
    Eq.~\eqref{eq:err}. The error is consistently around $\SI{1.5}{\percent}$. The case
    with mass ratio $29\slash 36$ is similar.}%
  \label{fig:final}
\end{figure}

In Fig.~\ref{fig:final}, we show the predictions of the model for mergers of
black holes with the same mass and charge-to-mass ratio and we plot the result
from the numerical-relativity simulations (squares).

We measure the total error as
\begin{equation}
  \label{eq:err}
  \text{RMS Error} = \sqrt{{\left(\frac{\chi_{\text{model}} - \chi_{\text {sim}}}{\chi_{\text{model}}}\right)}^{2}
  +  {\left(\frac{\lambda_{\text{model}} - \lambda_{\text{sim}}}{\lambda_{\text {model}}}\right)}^{2}}\,,
\end{equation}
and find that the error is about \SI{1.5}{\percent} independently of the value of $\lambda$
(bottom panel). By comparing the two terms in Eq.~\eqref{eq:err}, we find that
most of the error comes from the charge-to-mass ratio as opposed to the spin. We
also find the same error level and behavior in the unequal-mass cases we
consider here. Therefore, we conclude that the method described in the previous
sections is effective at predicting the properties of the remnant left by the
merger of two charged black holes with mass ratios close to unity, and equal
charge-to-mass ratio. Considering the complex and non-linear system under
consideration, this is a remarkable agreement.

Our numerical relativity simulations verified that the method described in
Section~\ref{sec:analytical-model} can correctly capture the properties of the
remnant left by the merger of charged black holes. With this, we can now look at
what happens when we consider the case with $\lambda \to 1$. In this, we are interested
in checking whether the remnant left by the merger would be over-extremal, i.e.,
$\lambda^{2} + \chi^{2} > 1$. Kerr-Newman spacetimes are the most general axisymmetric
and stationary four-dimensional electrovacuum spacetimes, and when
$\lambda^{2} + \chi^{2} > 1$ they describe naked singularities. So, a merger of charge
black holes such that the remnant has $\lambda^{2} + \chi^{2} > 1$ would be a good
candidate to violate cosmic censorship.

At this point, it is useful to recall what happens in the case of the merger of
two uncharged black holes with spin $\chi_{0} \to 1$. If we assumed that all the
angular momentum and mass end up in the remnant black hole, we would find that
the Kerr remnant has spin (for identical black holes with prograde dimensionless
spin $\chi_{0}$) $2\chi_{0} + \chi_{\text{orbital}} > 1$, hinting to a violation
of the cosmic censorship conjecture. This does not happen because the vast
majority of the total angular momentum is radiated away through emission of
gravitational waves.  In particular, the system orbits for longer than the
non-spinning case to radiate all the excess angular momentum and ensure that the
remnant is a black hole and not a naked singularity. We can construct a similar
thought experiment for the case of charge, with the difference that charge is
conserved. Therefore, if we start with charge-to-mass ratio $\lambda_{0}$, the
remnant must have $\lambda > 2 \lambda_{0}$ (because there is emission of
energy), and its spin must be greater than 0. So, it appears that there could be
conditions that favor $\lambda^{2} + \chi^{2} > 1$. Given charge conservation,
Nature has to find a new way to avoid the formation of a naked singularity, if
cosmic censorship is not violated.

\begin{figure}[htbp]
  \centering
  \includegraphics[width=\linewidth]{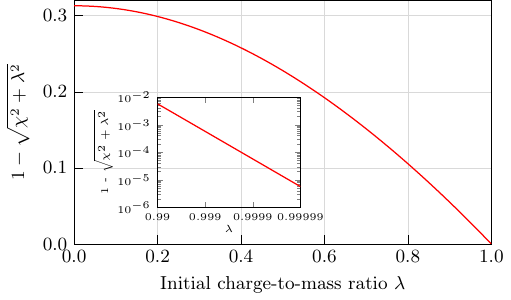}
  \caption{Kerr-Newman parameter as a function of the initial charge-to-mass
    ratio for quasi-circular mergers of equal-mass, equal-charge binaries, as
    predicted with the method described in this paper. The inset shows that even
    with $\lambda \to 1$, we have that the Kerr-Newman parameter is larger than
    0.}%
  \label{fig:kerr_newman_parameter}
\end{figure}

In Fig.~\ref{fig:kerr_newman_parameter}, we plot the prediction for $1 - \sqrt
{\lambda^{2} + \chi^{2}}$ for the remnant left by the merger of two equal mass,
equal charge binaries according to the analytical model described earlie.
Cosmic censorship demands that $1 - \sqrt {\lambda^{2} + \chi^{2}} > 0$ ($M^{2}
> Q^{2} + a^{2}$). The plot shows that it is possible to have a remnant that is
arbitrary close to extremality (with $\lambda^{2} + \chi^{2} \to 1$), but it is
not possible to pass this limit. %
This means that the quasi circular mergers of charged black holes should not be
expected to lead to naked singularities. The reason for this is different from
the case of purely spinning black holes, where is the emission of angular
momentum that prevents the over-extremality, and more similar to the results
found in~\cite{Zilhao2012} for head-on collisions. We find is that with higher
charge, the binary is less and less bound, and the orbital acceleration is
smaller and smaller. With smaller acceleration, there is weaker emission of
gravitational and electromagnetic waves. In the limit of $\lambda \to 1$, the
binary takes an infinite amount of time to merge, emitting a vanishing amount of
energy. For $\lambda$ identically equal to 1, the system is in equilibrium with
gravitational and electrostatic forces canceling out (this is the
Majumdar-Papapetrou solution~\cite{Majumdar1947, Papapetrou1945,
  Bozzola2019}). In our simulations we find that a binary with $\lambda=0.6$
orbits twice as many times as uncharged binary before merger. Moreover, with
increasing $\lambda$, the orbital angular momentum decreases (it roughly goes as
$\sqrt{1 - \lambda^2}$), so that the spin of the remnant becomes arbitrarily
small (as seen in the top panel of Fig.~\ref{fig:final}).

\section{Conclusions}%
\label{sec:conclusions}

In this paper, we presented an analytical and computationally cheap method to
estimate the properties of the remnant left by the merger of two charged black
holes. The method is an extension of the technique developed in
Refs.~\cite{Buonanno2008, Kesden2008} which uses an effective-one-body treatment
and the properties of the innermost-stable circular orbit in an equivalent
Kerr-Newman spacetime. We performed numerical relativity simulations and
verified that the method is accurate at the percent level. This shows that the
simple argument is remarkably effective at quantitative predictions for the
properties of the remnant.

While the results presented here are only for quasi-circular mergers, we expect
them to hold for eccentric as well, because these orbits are bound, too. The
analytical method presented here essentially compares the energy at infinity and
the energy at the ISCO, so it does not matter how one reaches the ISCO\@ (as in
the case without charge~\cite{Hinder2007}). The only exception is that for
highly eccentric mergers one would need to consider the ISCO for these
orbits. The analytic model we presented works for arbitrary mass, spin, and
charge configurations, but our validation only involved quasi-circular mergers
with non-spinning black holes of comparable mass and charge-to-mass ratio values
up to 0.6. When more numerical-relativity simulations of charged inspirals will
be available, the validation can be extended. This is one of the possible
limitations of the argument that overextremal black holes cannot form in
quasicircular mergers of charged black holes.

Our second goal was to learn more about quasi-circular mergers of highly charged
black holes in the context of the cosmic censorship conjecture. Charge is 
conserved quantity, whereas energy and angular momentum are not. When two
identical black holes with charge-to-mass ratio $\lambda_{0}$ merge, the remnant
has to have $\lambda > 2 \lambda_{0}$ because of emission of energy. Moreover,
the dimensionless spin $\chi$ of the remnant has to be larger than 0. This opens
up the possibility that some initial configurations might lead to a remnant with
$\lambda^{2} + \chi^{2} > 1$. Given that the only axisymmetric and stationary
spacetime is the Kerr-Newman one and that $\lambda^{2} + \chi^{2} > 1$ would
mean that there is no horizon, finding such a configuration would hint to a
possible way of forming naked singularities. In the case of spin, this is
avoided via emission of angular momentum and the orbital hang-up. Our study
shows that in the charged black hole case increasing the initial black hole
charge makes the system less and less dynamical. With $\lambda_{0} \to 1$, the
system asymptotically takes an infinite amount of time to merge and radiate a
vanishing amount of gravitational and electromagnetic waves.  Moreover, when
$\lambda_{0} \to 1$ the orbital angular momentum also vanishes.  Therefore, we
conclude that while it is possible to produce remnants that are arbitrarily
close to extremality, it is not possible to break the limit.

One of the reasons why it is not possible to form naked singularities is that
the system is bound, which sets a limit on the available orbital angular
momentum. In future studies, we will test the limits of our model by considering
the case of unlike charges (which increase the angular momentum). We will also
treat the case of unbound orbits, such as the zoom-and-whirl
ones~\cite{Sperhake2009}.

\begin{acknowledgments}
  \small We used \texttt{kuibit}~\cite{kuibit} for part of our analysis.
  \texttt{kuibit}~\cite{kuibit} use \texttt{NumPy}~\cite{NumPy},
  \texttt{h5py}~\cite{h5py}, \texttt{SciPy}~\cite{SciPy}, and
  \texttt{SymPy}~\cite{SymPy}. We are grateful to the developers and maintainers
  of the open-source codes that we used. We thank Cyrus Worley for his help with
  the eccentricity estimates in simulation data. This work was in part supported
  by NSF Grants PHY-1912619, and PHY-2145421, as well as NASA Grant
  80NSSC20K1542 to the University of Arizona. This work used \texttt{Expanse}
  (funded by the NSF through award OAC-1928224) at the San Diego Supercomputing
  Center (SDSC), through allocation TG-PHY190020 from the Advanced
  Cyberinfrastructure Coordination Ecosystem: Services \& Support (ACCESS)
  program, which is supported by National Science Foundation grants \#2138259,
  \#2138286, \#2138307, \#2137603, and \#2138296. The work also used NASA's High End
  Computing clusters.
\end{acknowledgments}

\bibliography{einsteintoolkit, binary_maximum_charged}

\def\prd{Phys. Rev. D}\def\prl{Phys. Rev. Lett.}\def\apjl{Astrophys. J.
  Lett.}\def\apjs{Astrophys. J. Suppl.}\def\apj{Astrophys. J.}\def\aj{Astron.
  J.}\def\aap{Astron. Astrophys.}\def\aaps{Astron. Astrophys.
  Suppl.}\def\araa{Ann. Rev. Astron. Astrophys.}\def\adp{Ann.
  Phy.}\def\cqg{Classical Quant. Grav.}\def\mnras{Mon. Not. R. Astron.
  Soc.}\def\physrep{Phys. Rep.}\def\nat{Nat.}\def\pasj{Pub. Astron. Soc. of
  Jap.}\def\jcap{J. Cosm. Astropart. Phys.}\def\apss{Astrophys. Spac. Sci.}
\begin{thebibliography}{45}
\expandafter\ifx\csname natexlab\endcsname\relax\def\natexlab#1{#1}\fi
\expandafter\ifx\csname bibnamefont\endcsname\relax
  \def\bibnamefont#1{#1}\fi
\expandafter\ifx\csname bibfnamefont\endcsname\relax
  \def\bibfnamefont#1{#1}\fi
\expandafter\ifx\csname citenamefont\endcsname\relax
  \def\citenamefont#1{#1}\fi
\expandafter\ifx\csname url\endcsname\relax
  \def\url#1{\texttt{#1}}\fi
\expandafter\ifx\csname urlprefix\endcsname\relax\def\urlprefix{URL }\fi
\providecommand{\bibinfo}[2]{#2}
\providecommand{\eprint}[2][]{\url{#2}}

\bibitem[{\citenamefont{{Campanelli} et~al.}(2006)\citenamefont{{Campanelli},
  {Lousto}, and {Zlochower}}}]{Campanelli2006}
\bibinfo{author}{\bibfnamefont{M.}~\bibnamefont{{Campanelli}}},
  \bibinfo{author}{\bibfnamefont{C.~O.} \bibnamefont{{Lousto}}},
  \bibnamefont{and}
  \bibinfo{author}{\bibfnamefont{Y.}~\bibnamefont{{Zlochower}}},
  \bibinfo{journal}{\prd} \textbf{\bibinfo{volume}{74}}, \bibinfo{eid}{041501}
  (\bibinfo{year}{2006}), \eprint{gr-qc/0604012}.

\bibitem[{\citenamefont{Wald}(1984)}]{Wald1984}
\bibinfo{author}{\bibfnamefont{R.~M.} \bibnamefont{Wald}},
  \emph{\bibinfo{title}{{General relativity}}} (\bibinfo{publisher}{Chicago
  Univ. Press}, \bibinfo{address}{Chicago, IL}, \bibinfo{year}{1984}),
  \urlprefix\url{https://cds.cern.ch/record/106274}.

\bibitem[{\citenamefont{{Choptuik}}(1993)}]{Choptuik1993}
\bibinfo{author}{\bibfnamefont{M.~W.} \bibnamefont{{Choptuik}}},
  \bibinfo{journal}{\prl} \textbf{\bibinfo{volume}{70}}, \bibinfo{pages}{9}
  (\bibinfo{year}{1993}).

\bibitem[{\citenamefont{Bozzola}(2022)}]{Bozzola2022}
\bibinfo{author}{\bibfnamefont{G.}~\bibnamefont{Bozzola}},
  \bibinfo{journal}{Phys. Rev. Lett.} \textbf{\bibinfo{volume}{128}},
  \bibinfo{pages}{071101} (\bibinfo{year}{2022}), \eprint{2202.05310}.

\bibitem[{\citenamefont{{Buonanno} et~al.}(2008)\citenamefont{{Buonanno},
  {Kidder}, and {Lehner}}}]{Buonanno2008}
\bibinfo{author}{\bibfnamefont{A.}~\bibnamefont{{Buonanno}}},
  \bibinfo{author}{\bibfnamefont{L.~E.} \bibnamefont{{Kidder}}},
  \bibnamefont{and} \bibinfo{author}{\bibfnamefont{L.}~\bibnamefont{{Lehner}}},
  \bibinfo{journal}{\prd} \textbf{\bibinfo{volume}{77}}, \bibinfo{eid}{026004}
  (\bibinfo{year}{2008}), \eprint{0709.3839}.

\bibitem[{\citenamefont{{Kesden}}(2008)}]{Kesden2008}
\bibinfo{author}{\bibfnamefont{M.}~\bibnamefont{{Kesden}}},
  \bibinfo{journal}{\prd} \textbf{\bibinfo{volume}{78}}, \bibinfo{eid}{084030}
  (\bibinfo{year}{2008}), \eprint{0807.3043}.

\bibitem[{\citenamefont{{Wald}}(2018)}]{Wald2018}
\bibinfo{author}{\bibfnamefont{R.~M.} \bibnamefont{{Wald}}},
  \bibinfo{journal}{International Journal of Modern Physics D}
  \textbf{\bibinfo{volume}{27}}, \bibinfo{eid}{1843003} (\bibinfo{year}{2018}).

\bibitem[{\citenamefont{{Bozzola} and {Paschalidis}}(2019)}]{Bozzola2019}
\bibinfo{author}{\bibfnamefont{G.}~\bibnamefont{{Bozzola}}} \bibnamefont{and}
  \bibinfo{author}{\bibfnamefont{V.}~\bibnamefont{{Paschalidis}}},
  \bibinfo{journal}{\prd} \textbf{\bibinfo{volume}{99}}, \bibinfo{eid}{104044}
  (\bibinfo{year}{2019}), \eprint{1903.01036}.

\bibitem[{\citenamefont{{Liu} et~al.}(2020)\citenamefont{{Liu}, {Guo}, {Cai},
  and {Kim}}}]{Liu2020}
\bibinfo{author}{\bibfnamefont{L.}~\bibnamefont{{Liu}}},
  \bibinfo{author}{\bibfnamefont{Z.-K.} \bibnamefont{{Guo}}},
  \bibinfo{author}{\bibfnamefont{R.-G.} \bibnamefont{{Cai}}}, \bibnamefont{and}
  \bibinfo{author}{\bibfnamefont{S.~P.} \bibnamefont{{Kim}}},
  \bibinfo{journal}{\prd} \textbf{\bibinfo{volume}{102}}, \bibinfo{eid}{043508}
  (\bibinfo{year}{2020}), \eprint{2001.02984}.

\bibitem[{\citenamefont{{Carullo} et~al.}(2021)\citenamefont{{Carullo},
  {Laghi}, {Johnson-McDaniel}, {Del Pozzo}, {Dias}, {Godazgar}, and
  {Santos}}}]{Carullo2021}
\bibinfo{author}{\bibfnamefont{G.}~\bibnamefont{{Carullo}}},
  \bibinfo{author}{\bibfnamefont{D.}~\bibnamefont{{Laghi}}},
  \bibinfo{author}{\bibfnamefont{N.~K.} \bibnamefont{{Johnson-McDaniel}}},
  \bibinfo{author}{\bibfnamefont{W.}~\bibnamefont{{Del Pozzo}}},
  \bibinfo{author}{\bibfnamefont{O.~J.~C.} \bibnamefont{{Dias}}},
  \bibinfo{author}{\bibfnamefont{M.}~\bibnamefont{{Godazgar}}},
  \bibnamefont{and} \bibinfo{author}{\bibfnamefont{J.~E.}
  \bibnamefont{{Santos}}}, \bibinfo{journal}{arXiv e-prints}
  \bibinfo{eid}{arXiv:2109.13961} (\bibinfo{year}{2021}), \eprint{2109.13961}.

\bibitem[{\citenamefont{{Hughes} and {Blandford}}(2003)}]{Hughes2003}
\bibinfo{author}{\bibfnamefont{S.~A.} \bibnamefont{{Hughes}}} \bibnamefont{and}
  \bibinfo{author}{\bibfnamefont{R.~D.} \bibnamefont{{Blandford}}},
  \bibinfo{journal}{\apjl} \textbf{\bibinfo{volume}{585}},
  \bibinfo{pages}{L101} (\bibinfo{year}{2003}), \eprint{astro-ph/0208484}.

\bibitem[{\citenamefont{Levenberg}(1944)}]{Levenberg1944}
\bibinfo{author}{\bibfnamefont{K.}~\bibnamefont{Levenberg}},
  \bibinfo{journal}{Quarterly of Applied Mathematics}
  \textbf{\bibinfo{volume}{2}}, \bibinfo{pages}{164} (\bibinfo{year}{1944}).

\bibitem[{\citenamefont{Marquardt}(1963)}]{Marquardt1963}
\bibinfo{author}{\bibfnamefont{D.~W.} \bibnamefont{Marquardt}},
  \bibinfo{journal}{Journal of the Society for Industrial and Applied
  Mathematics} \textbf{\bibinfo{volume}{11}}, \bibinfo{pages}{431}
  (\bibinfo{year}{1963}).

\bibitem[{\citenamefont{{Virtanen} et~al.}(2020)\citenamefont{{Virtanen},
  {Gommers}, {Oliphant}, {Haberland}, {Reddy}, {Cournapeau}, {Burovski},
  {Peterson}, {Weckesser}, {Bright} et~al.}}]{SciPy}
\bibinfo{author}{\bibfnamefont{P.}~\bibnamefont{{Virtanen}}},
  \bibinfo{author}{\bibfnamefont{R.}~\bibnamefont{{Gommers}}},
  \bibinfo{author}{\bibfnamefont{T.~E.} \bibnamefont{{Oliphant}}},
  \bibinfo{author}{\bibfnamefont{M.}~\bibnamefont{{Haberland}}},
  \bibinfo{author}{\bibfnamefont{T.}~\bibnamefont{{Reddy}}},
  \bibinfo{author}{\bibfnamefont{D.}~\bibnamefont{{Cournapeau}}},
  \bibinfo{author}{\bibfnamefont{E.}~\bibnamefont{{Burovski}}},
  \bibinfo{author}{\bibfnamefont{P.}~\bibnamefont{{Peterson}}},
  \bibinfo{author}{\bibfnamefont{W.}~\bibnamefont{{Weckesser}}},
  \bibinfo{author}{\bibfnamefont{J.}~\bibnamefont{{Bright}}},
  \bibnamefont{et~al.}, \bibinfo{journal}{Nature Methods}
  \textbf{\bibinfo{volume}{17}}, \bibinfo{pages}{261} (\bibinfo{year}{2020}).

\bibitem[{\citenamefont{{Jaiakson} et~al.}(2017)\citenamefont{{Jaiakson},
  {Chatrabhuti}, {Evnin}, and {Lehner}}}]{Jaiakson2017}
\bibinfo{author}{\bibfnamefont{P.}~\bibnamefont{{Jaiakson}}},
  \bibinfo{author}{\bibfnamefont{A.}~\bibnamefont{{Chatrabhuti}}},
  \bibinfo{author}{\bibfnamefont{O.}~\bibnamefont{{Evnin}}}, \bibnamefont{and}
  \bibinfo{author}{\bibfnamefont{L.}~\bibnamefont{{Lehner}}},
  \bibinfo{journal}{\prd} \textbf{\bibinfo{volume}{96}}, \bibinfo{eid}{044031}
  (\bibinfo{year}{2017}), \eprint{1706.06519}.

\bibitem[{\citenamefont{{Siahaan}}(2020)}]{Siahaan2020}
\bibinfo{author}{\bibfnamefont{H.~M.} \bibnamefont{{Siahaan}}},
  \bibinfo{journal}{\prd} \textbf{\bibinfo{volume}{101}}, \bibinfo{eid}{064036}
  (\bibinfo{year}{2020}), \eprint{1907.02158}.

\bibitem[{\citenamefont{{Bozzola} and {Paschalidis}}(2021)}]{Bozzola2021}
\bibinfo{author}{\bibfnamefont{G.}~\bibnamefont{{Bozzola}}} \bibnamefont{and}
  \bibinfo{author}{\bibfnamefont{V.}~\bibnamefont{{Paschalidis}}},
  \bibinfo{journal}{\prd} \textbf{\bibinfo{volume}{104}}, \bibinfo{eid}{044004}
  (\bibinfo{year}{2021}), \eprint{2104.06978}.

\bibitem[{\citenamefont{Meurer et~al.}(2017)\citenamefont{Meurer, Smith,
  Paprocki, \v{C}ert\'{i}k, Kirpichev, Rocklin, Kumar, Ivanov, Moore, Singh
  et~al.}}]{SymPy}
\bibinfo{author}{\bibfnamefont{A.}~\bibnamefont{Meurer}},
  \bibinfo{author}{\bibfnamefont{C.~P.} \bibnamefont{Smith}},
  \bibinfo{author}{\bibfnamefont{M.}~\bibnamefont{Paprocki}},
  \bibinfo{author}{\bibfnamefont{O.}~\bibnamefont{\v{C}ert\'{i}k}},
  \bibinfo{author}{\bibfnamefont{S.~B.} \bibnamefont{Kirpichev}},
  \bibinfo{author}{\bibfnamefont{M.}~\bibnamefont{Rocklin}},
  \bibinfo{author}{\bibfnamefont{A.}~\bibnamefont{Kumar}},
  \bibinfo{author}{\bibfnamefont{S.}~\bibnamefont{Ivanov}},
  \bibinfo{author}{\bibfnamefont{J.~K.} \bibnamefont{Moore}},
  \bibinfo{author}{\bibfnamefont{S.}~\bibnamefont{Singh}},
  \bibnamefont{et~al.}, \bibinfo{journal}{PeerJ Computer Science}
  \textbf{\bibinfo{volume}{3}}, \bibinfo{pages}{e103} (\bibinfo{year}{2017}),
  ISSN \bibinfo{issn}{2376-5992},
  \urlprefix\url{https://doi.org/10.7717/peerj-cs.103}.

\bibitem[{\citenamefont{Bozzola and Paschalidis}(2021)}]{Bozzola2020}
\bibinfo{author}{\bibfnamefont{G.}~\bibnamefont{Bozzola}} \bibnamefont{and}
  \bibinfo{author}{\bibfnamefont{V.}~\bibnamefont{Paschalidis}},
  \bibinfo{journal}{Phys. Rev. Lett.} \textbf{\bibinfo{volume}{126}},
  \bibinfo{pages}{041103} (\bibinfo{year}{2021}), \eprint{2006.15764}.

\bibitem[{\citenamefont{Luna et~al.}(2022)\citenamefont{Luna, Bozzola, Cardoso,
  Paschalidis, and Zilh\~ao}}]{Luna:2022udb}
\bibinfo{author}{\bibfnamefont{R.}~\bibnamefont{Luna}},
  \bibinfo{author}{\bibfnamefont{G.}~\bibnamefont{Bozzola}},
  \bibinfo{author}{\bibfnamefont{V.}~\bibnamefont{Cardoso}},
  \bibinfo{author}{\bibfnamefont{V.}~\bibnamefont{Paschalidis}},
  \bibnamefont{and} \bibinfo{author}{\bibfnamefont{M.}~\bibnamefont{Zilh\~ao}},
  \bibinfo{journal}{Phys. Rev. D} \textbf{\bibinfo{volume}{106}},
  \bibinfo{pages}{084017} (\bibinfo{year}{2022}), \eprint{2207.06429}.

\bibitem[{\citenamefont{{Alcubierre}}(2008)}]{Alcubierre:2008it}
\bibinfo{author}{\bibfnamefont{M.}~\bibnamefont{{Alcubierre}}},
  \emph{\bibinfo{title}{{Introduction to 3+1 Numerical Relativity}}}
  (\bibinfo{publisher}{Oxford University Press, UK}, \bibinfo{year}{2008}),
  ISBN \bibinfo{isbn}{978-0-19-920567-7}.

\bibitem[{\citenamefont{{Baumgarte} and {Shapiro}}(2010)}]{Baumgarte:2010nu}
\bibinfo{author}{\bibfnamefont{T.~W.} \bibnamefont{{Baumgarte}}}
  \bibnamefont{and} \bibinfo{author}{\bibfnamefont{S.~L.}
  \bibnamefont{{Shapiro}}}, \emph{\bibinfo{title}{{Numerical Relativity:
  Solving Einstein's Equations on the Computer}}}
  (\bibinfo{publisher}{Cambridge University Press}, \bibinfo{year}{2010}), ISBN
  \bibinfo{isbn}{978-0-52-151407-1}.

\bibitem[{\citenamefont{{Shibata}}(2016)}]{Shibata2016b}
\bibinfo{author}{\bibfnamefont{M.}~\bibnamefont{{Shibata}}},
  \emph{\bibinfo{title}{{Numerical Relativity}}} (\bibinfo{publisher}{World
  Scientific Publishing Co}, \bibinfo{year}{2016}).

\bibitem[{\citenamefont{L{\"{o}}ffler et~al.}(2012)\citenamefont{L{\"{o}}ffler,
  Faber, Bentivegna, Bode, Diener, Haas, Hinder, Mundim, Ott, Schnetter
  et~al.}}]{Loffler:2011ay}
\bibinfo{author}{\bibfnamefont{F.}~\bibnamefont{L{\"{o}}ffler}},
  \bibinfo{author}{\bibfnamefont{J.}~\bibnamefont{Faber}},
  \bibinfo{author}{\bibfnamefont{E.}~\bibnamefont{Bentivegna}},
  \bibinfo{author}{\bibfnamefont{T.}~\bibnamefont{Bode}},
  \bibinfo{author}{\bibfnamefont{P.}~\bibnamefont{Diener}},
  \bibinfo{author}{\bibfnamefont{R.}~\bibnamefont{Haas}},
  \bibinfo{author}{\bibfnamefont{I.}~\bibnamefont{Hinder}},
  \bibinfo{author}{\bibfnamefont{B.~C.} \bibnamefont{Mundim}},
  \bibinfo{author}{\bibfnamefont{C.~D.} \bibnamefont{Ott}},
  \bibinfo{author}{\bibfnamefont{E.}~\bibnamefont{Schnetter}},
  \bibnamefont{et~al.}, \bibinfo{journal}{Class. Quantum Grav.}
  \textbf{\bibinfo{volume}{29}}, \bibinfo{pages}{115001}
  (\bibinfo{year}{2012}), \eprint{arXiv:1111.3344 [gr-qc]}.

\bibitem[{\citenamefont{{Collaborative Effort}}(2011)}]{EinsteinToolkit:ascl}
\bibinfo{author}{\bibnamefont{{Collaborative Effort}}},
  \emph{\bibinfo{title}{{Einstein Toolkit for Relativistic Astrophysics}}},
  \bibinfo{howpublished}{Astrophysics Source Code Library}
  (\bibinfo{year}{2011}), \eprint{ascl:1102.014}.

\bibitem[{EinsteinToolkit()}]{EinsteinToolkit:web}
EinsteinToolkit, \emph{\bibinfo{title}{{Einstein Toolkit}: Open software for
  relativistic astrophysics}}, \urlprefix\url{http://einsteintoolkit.org/}.

\bibitem[{\citenamefont{Brandt et~al.}(2021)\citenamefont{Brandt, Bozzola,
  Cheng, Diener, Dima, Gabella, Gracia-Linares, Haas, Zlochower, Alcubierre
  et~al.}}]{EinsteinToolkit:2021_11}
\bibinfo{author}{\bibfnamefont{S.~R.} \bibnamefont{Brandt}},
  \bibinfo{author}{\bibfnamefont{G.}~\bibnamefont{Bozzola}},
  \bibinfo{author}{\bibfnamefont{C.-H.} \bibnamefont{Cheng}},
  \bibinfo{author}{\bibfnamefont{P.}~\bibnamefont{Diener}},
  \bibinfo{author}{\bibfnamefont{A.}~\bibnamefont{Dima}},
  \bibinfo{author}{\bibfnamefont{W.~E.} \bibnamefont{Gabella}},
  \bibinfo{author}{\bibfnamefont{M.}~\bibnamefont{Gracia-Linares}},
  \bibinfo{author}{\bibfnamefont{R.}~\bibnamefont{Haas}},
  \bibinfo{author}{\bibfnamefont{Y.}~\bibnamefont{Zlochower}},
  \bibinfo{author}{\bibfnamefont{M.}~\bibnamefont{Alcubierre}},
  \bibnamefont{et~al.}, \emph{\bibinfo{title}{The einstein toolkit}}
  (\bibinfo{year}{2021}), \bibinfo{note}{to find out more, visit
  http://einsteintoolkit.org},
  \urlprefix\url{https://doi.org/10.5281/zenodo.5770803}.

\bibitem[{\citenamefont{{Sperhake}}(2007)}]{Sperhake2007}
\bibinfo{author}{\bibfnamefont{U.}~\bibnamefont{{Sperhake}}},
  \bibinfo{journal}{\prd} \textbf{\bibinfo{volume}{76}}, \bibinfo{eid}{104015}
  (\bibinfo{year}{2007}), \eprint{gr-qc/0606079}.

\bibitem[{\citenamefont{Shibata and Nakamura}(1995)}]{Shibata1995}
\bibinfo{author}{\bibfnamefont{M.}~\bibnamefont{Shibata}} \bibnamefont{and}
  \bibinfo{author}{\bibfnamefont{T.}~\bibnamefont{Nakamura}},
  \bibinfo{journal}{Phys. Rev. D} \textbf{\bibinfo{volume}{52}},
  \bibinfo{pages}{5428} (\bibinfo{year}{1995}),
  \urlprefix\url{https://link.aps.org/doi/10.1103/PhysRevD.52.5428}.

\bibitem[{\citenamefont{Baumgarte and Shapiro}(1998)}]{Baumgarte1998}
\bibinfo{author}{\bibfnamefont{T.~W.} \bibnamefont{Baumgarte}}
  \bibnamefont{and} \bibinfo{author}{\bibfnamefont{S.~L.}
  \bibnamefont{Shapiro}}, \bibinfo{journal}{Phys. Rev. D}
  \textbf{\bibinfo{volume}{59}}, \bibinfo{pages}{024007}
  (\bibinfo{year}{1998}).

\bibitem[{\citenamefont{{Zilh{\~a}o} et~al.}(2015)\citenamefont{{Zilh{\~a}o},
  {Witek}, and {Cardoso}}}]{Zilhao2015}
\bibinfo{author}{\bibfnamefont{M.}~\bibnamefont{{Zilh{\~a}o}}},
  \bibinfo{author}{\bibfnamefont{H.}~\bibnamefont{{Witek}}}, \bibnamefont{and}
  \bibinfo{author}{\bibfnamefont{V.}~\bibnamefont{{Cardoso}}},
  \bibinfo{journal}{\cqg} \textbf{\bibinfo{volume}{32}}, \bibinfo{eid}{234003}
  (\bibinfo{year}{2015}), \eprint{1505.00797}.

\bibitem[{\citenamefont{Witek et~al.}(2021)\citenamefont{Witek, Zilhao,
  Bozzola, Elley, Ficarra, Ikeda, Sanchis-Gual, and Silva}}]{canuda}
\bibinfo{author}{\bibfnamefont{H.}~\bibnamefont{Witek}},
  \bibinfo{author}{\bibfnamefont{M.}~\bibnamefont{Zilhao}},
  \bibinfo{author}{\bibfnamefont{G.}~\bibnamefont{Bozzola}},
  \bibinfo{author}{\bibfnamefont{M.}~\bibnamefont{Elley}},
  \bibinfo{author}{\bibfnamefont{G.}~\bibnamefont{Ficarra}},
  \bibinfo{author}{\bibfnamefont{T.}~\bibnamefont{Ikeda}},
  \bibinfo{author}{\bibfnamefont{N.}~\bibnamefont{Sanchis-Gual}},
  \bibnamefont{and} \bibinfo{author}{\bibfnamefont{H.}~\bibnamefont{Silva}},
  \emph{\bibinfo{title}{{Canuda: a public numerical relativity library to probe
  fundamental physics}}} (\bibinfo{year}{2021}),
  \urlprefix\url{https://doi.org/10.5281/zenodo.5520862}.

\bibitem[{\citenamefont{{Witek} and {Zilh{\~a}o}}(2015)}]{canudacode}
\bibinfo{author}{\bibfnamefont{H.}~\bibnamefont{{Witek}}} \bibnamefont{and}
  \bibinfo{author}{\bibfnamefont{M.}~\bibnamefont{{Zilh{\~a}o}}},
  \emph{\bibinfo{title}{Canuda code}} (\bibinfo{year}{2015}),
  \urlprefix\url{https://bitbucket.org/canuda/}.

\bibitem[{\citenamefont{Thornburg}(1996)}]{Thornburg:1995cp}
\bibinfo{author}{\bibfnamefont{J.}~\bibnamefont{Thornburg}},
  \bibinfo{journal}{Phys. Rev. D} \textbf{\bibinfo{volume}{54}},
  \bibinfo{pages}{4899} (\bibinfo{year}{1996}), \eprint{arXiv:gr-qc/9508014}.

\bibitem[{\citenamefont{Thornburg}(2004)}]{Thornburg:2003sf}
\bibinfo{author}{\bibfnamefont{J.}~\bibnamefont{Thornburg}},
  \bibinfo{journal}{Class. Quantum Grav.} \textbf{\bibinfo{volume}{21}},
  \bibinfo{pages}{743} (\bibinfo{year}{2004}), \eprint{arXiv:gr-qc/0306056}.

\bibitem[{\citenamefont{Dreyer et~al.}(2003)\citenamefont{Dreyer, Krishnan,
  Shoemaker, and Schnetter}}]{Dreyer:2002mx}
\bibinfo{author}{\bibfnamefont{O.}~\bibnamefont{Dreyer}},
  \bibinfo{author}{\bibfnamefont{B.}~\bibnamefont{Krishnan}},
  \bibinfo{author}{\bibfnamefont{D.}~\bibnamefont{Shoemaker}},
  \bibnamefont{and}
  \bibinfo{author}{\bibfnamefont{E.}~\bibnamefont{Schnetter}},
  \bibinfo{journal}{Phys. Rev. D} \textbf{\bibinfo{volume}{67}},
  \bibinfo{pages}{024018} (\bibinfo{year}{2003}), \eprint{arXiv:gr-qc/0206008}.

\bibitem[{\citenamefont{Schnetter et~al.}(2004)\citenamefont{Schnetter, Hawley,
  and Hawke}}]{Schnetter:2003rb}
\bibinfo{author}{\bibfnamefont{E.}~\bibnamefont{Schnetter}},
  \bibinfo{author}{\bibfnamefont{S.~H.} \bibnamefont{Hawley}},
  \bibnamefont{and} \bibinfo{author}{\bibfnamefont{I.}~\bibnamefont{Hawke}},
  \bibinfo{journal}{Class. Quantum Grav.} \textbf{\bibinfo{volume}{21}},
  \bibinfo{pages}{1465} (\bibinfo{year}{2004}), \eprint{arXiv:gr-qc/0310042}.

\bibitem[{\citenamefont{{Zilh{\~a}o} et~al.}(2012)\citenamefont{{Zilh{\~a}o},
  {Cardoso}, {Herdeiro}, {Lehner}, and {Sperhake}}}]{Zilhao2012}
\bibinfo{author}{\bibfnamefont{M.}~\bibnamefont{{Zilh{\~a}o}}},
  \bibinfo{author}{\bibfnamefont{V.}~\bibnamefont{{Cardoso}}},
  \bibinfo{author}{\bibfnamefont{C.}~\bibnamefont{{Herdeiro}}},
  \bibinfo{author}{\bibfnamefont{L.}~\bibnamefont{{Lehner}}}, \bibnamefont{and}
  \bibinfo{author}{\bibfnamefont{U.}~\bibnamefont{{Sperhake}}},
  \bibinfo{journal}{\prd} \textbf{\bibinfo{volume}{85}}, \bibinfo{eid}{124062}
  (\bibinfo{year}{2012}), \eprint{1205.1063}.

\bibitem[{\citenamefont{{Majumdar}}(1947)}]{Majumdar1947}
\bibinfo{author}{\bibfnamefont{S.~D.} \bibnamefont{{Majumdar}}},
  \bibinfo{journal}{Physical Review} \textbf{\bibinfo{volume}{72}},
  \bibinfo{pages}{390} (\bibinfo{year}{1947}).

\bibitem[{\citenamefont{Papapetrou}(1945)}]{Papapetrou1945}
\bibinfo{author}{\bibfnamefont{A.}~\bibnamefont{Papapetrou}},
  \bibinfo{journal}{Proceedings of the Royal Irish Academy. Section A:
  Mathematical and Physical Sciences} \textbf{\bibinfo{volume}{51}},
  \bibinfo{pages}{191} (\bibinfo{year}{1945}), ISSN \bibinfo{issn}{00358975},
  \urlprefix\url{http://www.jstor.org/stable/20488481}.

\bibitem[{\citenamefont{Hinder et~al.}(2008)\citenamefont{Hinder, Vaishnav,
  Herrmann, Shoemaker, and Laguna}}]{Hinder2007}
\bibinfo{author}{\bibfnamefont{I.}~\bibnamefont{Hinder}},
  \bibinfo{author}{\bibfnamefont{B.}~\bibnamefont{Vaishnav}},
  \bibinfo{author}{\bibfnamefont{F.}~\bibnamefont{Herrmann}},
  \bibinfo{author}{\bibfnamefont{D.}~\bibnamefont{Shoemaker}},
  \bibnamefont{and} \bibinfo{author}{\bibfnamefont{P.}~\bibnamefont{Laguna}},
  \bibinfo{journal}{Phys. Rev. D} \textbf{\bibinfo{volume}{77}},
  \bibinfo{pages}{081502} (\bibinfo{year}{2008}), \eprint{0710.5167}.

\bibitem[{\citenamefont{{Sperhake} et~al.}(2009)\citenamefont{{Sperhake},
  {Cardoso}, {Pretorius}, {Berti}, {Hinderer}, and {Yunes}}}]{Sperhake2009}
\bibinfo{author}{\bibfnamefont{U.}~\bibnamefont{{Sperhake}}},
  \bibinfo{author}{\bibfnamefont{V.}~\bibnamefont{{Cardoso}}},
  \bibinfo{author}{\bibfnamefont{F.}~\bibnamefont{{Pretorius}}},
  \bibinfo{author}{\bibfnamefont{E.}~\bibnamefont{{Berti}}},
  \bibinfo{author}{\bibfnamefont{T.}~\bibnamefont{{Hinderer}}},
  \bibnamefont{and} \bibinfo{author}{\bibfnamefont{N.}~\bibnamefont{{Yunes}}},
  \bibinfo{journal}{Physical Review Letters} \textbf{\bibinfo{volume}{103}},
  \bibinfo{eid}{131102} (\bibinfo{year}{2009}), \eprint{0907.1252}.

\bibitem[{\citenamefont{Bozzola}(2021)}]{kuibit}
\bibinfo{author}{\bibfnamefont{G.}~\bibnamefont{Bozzola}},
  \bibinfo{journal}{The Journal of Open Source Software}
  \textbf{\bibinfo{volume}{6}}, \bibinfo{pages}{3099} (\bibinfo{year}{2021}),
  \eprint{2104.06376}.

\bibitem[{\citenamefont{Harris et~al.}(2020)\citenamefont{Harris, Millman,
  van~der Walt, Gommers, Virtanen, Cournapeau, Wieser, Taylor, Berg, Smith
  et~al.}}]{NumPy}
\bibinfo{author}{\bibfnamefont{C.~R.} \bibnamefont{Harris}},
  \bibinfo{author}{\bibfnamefont{K.~J.} \bibnamefont{Millman}},
  \bibinfo{author}{\bibfnamefont{S.~J.} \bibnamefont{van~der Walt}},
  \bibinfo{author}{\bibfnamefont{R.}~\bibnamefont{Gommers}},
  \bibinfo{author}{\bibfnamefont{P.}~\bibnamefont{Virtanen}},
  \bibinfo{author}{\bibfnamefont{D.}~\bibnamefont{Cournapeau}},
  \bibinfo{author}{\bibfnamefont{E.}~\bibnamefont{Wieser}},
  \bibinfo{author}{\bibfnamefont{J.}~\bibnamefont{Taylor}},
  \bibinfo{author}{\bibfnamefont{S.}~\bibnamefont{Berg}},
  \bibinfo{author}{\bibfnamefont{N.~J.} \bibnamefont{Smith}},
  \bibnamefont{et~al.}, \bibinfo{journal}{Nature}
  \textbf{\bibinfo{volume}{585}}, \bibinfo{pages}{357} (\bibinfo{year}{2020}),
  \urlprefix\url{https://doi.org/10.1038/s41586-020-2649-2}.

\bibitem[{\citenamefont{Collette}(2013)}]{h5py}
\bibinfo{author}{\bibfnamefont{A.}~\bibnamefont{Collette}},
  \emph{\bibinfo{title}{Python and HDF5}} (\bibinfo{publisher}{O'Reilly},
  \bibinfo{year}{2013}).

\end{thebibliography}

\end{document}